\documentclass[num-refs]{wiley-article}
\usepackage{jf-names}
\usepackage[utf8]{inputenc}
\usepackage{graphicx}
\usepackage{color}
\usepackage{caption}
\usepackage{subcaption}
\usepackage{amsmath}
\usepackage{amssymb}

\papertype{Full Paper}
\paperfield{Magnetic Resonance in Medicine}

\title{Advancing machine learning for MR image reconstruction with an open competition: Overview of the 2019 fastMRI challenge}

\author[1]{Florian~Knoll$^\dagger$}
\author[2]{Tullie~Murrell$^\dagger$}
\author[2]{Anuroop~Sriram$^\dagger$}
\author[2]{Nafissa~Yakubova}
\author[2]{Jure~Zbontar}
\author[2]{Michael~Rabbat}
\author[2]{Aaron~Defazio}
\author[1]{Matthew~J.~Muckley}
\author[1]{Daniel~K.~Sodickson}
\author[2]{C.~Lawrence~Zitnick}
\author[1]{Michael~P.~Recht}

\date{December 2019}

\affil[1]{Center for Advanced Imaging Innovation and Research (CAI$^2$R), Department of Radiology, New York University Grossman School of Medicine, New York, NY, 10016 United States}
\affil[2]{Facebook AI Research, Menlo Park, CA, 94025 United States}
\affil[$\dagger$]{Indicates equal contributions.}

\corraddress{Florian Knoll, Department of Radiology, NYU Grossman School of Medicine, New York, NY, 10016, United States}
\corremail{florian.knoll@nyumc.org}

\fundinginfo{
NIH: NIBIB, Awards NIH R01EB024532 and NIH P41EB017183.
\bigskip

\textbf{Submitted to Magnetic Resonance in Medicine}}

\runningauthor{Knoll et al.}

\begin{document}

\maketitle

\begin{abstract}
\textbf{Purpose:} To advance research in the field of machine learning for MR image reconstruction with an open challenge.

\noindent \textbf{Methods:} We provided participants with a dataset of raw k-space data from 1,594 consecutive clinical exams of the knee. The goal of the challenge was to reconstruct images from these data. In order to strike a balance between realistic data and a shallow learning curve for those not already familiar with MR image reconstruction, we ran multiple tracks for multi-coil and single-coil data. We performed a two-stage evaluation based on quantitative image metrics followed by evaluation by a panel of radiologists. The challenge ran from June to December of 2019.

\noindent \textbf{Results:} We received a total of 33 challenge submissions. All participants chose to submit results from supervised machine learning approaches.

\noindent \textbf{Conclusion:} The challenge led to new developments in machine learning for image reconstruction, provided insight into the current state of the art in the field, and highlighted remaining hurdles for clinical adoption.

\keywords{Challenge, Image reconstruction, Parallel imaging, Machine Learning, Compressed Sensing, Fast Imaging, Optimization, Public Dataset}
\end{abstract}

\section{Introduction}
One of the fastest growing fields of research in medical imaging during the last several years is the use of machine learning methods for image reconstruction. Machine learning has been proposed for CT dose reduction~\cite{Kang2017,Chen2017a,jin2017,Wolterink2017,Kobler2017,Adler2018}, attenuation correction for PET-MRI~\cite{Liu2018} and accelerated MR imaging~\cite{Hammernik2016,Wang2016,Hammernik2017,Aggarwal2019,Ye2017,Schlemper2018,Qin2019,Zhu2018,Chen2018}. Despite various methodological advances, the methods developed in these studies were all trained and validated on small individual datasets collected by the authors, which in many cases were not shared with the research community. These limitations in data accessibility makes it challenging to reproduce different approaches, and to validate comparisons between them. The lack of broadly accessible data also restricts work on important medical image reconstruction problems to researchers associated with or cooperating with large university medical centers where imaging data is available. This restriction is a significant lost opportunity, given the substantial volume of ongoing research in basic science machine learning and data science. 

Indeed, there is a striking contrast between specialized medical research and more general research in the field of machine learning, which has seen breakthrough improvements in diverse areas from image classification~\cite{Krizhevsky2012} with deep convolutional neural networks (CNNs) ~\cite{LeCun2015} to championship-level gaming~\cite{Silver2016}. The core technologies that led to these results had already been introduced around 1990 for applications like speech recognition~\cite{Waibel1989} and written document parsing~\cite{LeCun1989}. However, deep learning for computer vision did not expand beyond simple digit recognition tasks for the next 20 years. In retrospect, a single event is often identified as the key catalyst for the recent resurgence of machine learning technology~\cite{LeCun2015}: The 2012 ImageNet Large Scale Visual Recognition Challenge (ILSVRC)~\cite{Russakovsky2015}, in which a deep CNN achieved spectacular results for an image classification task~\cite{Krizhevsky2012}. Since then, every single winning entry in the competition was a form of deep CNN, with current winning entries even outperforming human performance. Winning the ILSVRC has become extremely prestigious and has attracted the interest of leading academic institutions and IT companies around the world. Performance on ILSVRC tasks has become a standard for the evaluation of new developments in computer vision. A similar event occurred in the field of medical image reconstruction with the 2016 Low Dose CT Grand Challenge organized by the Mayo Clinic~\cite{McCollough2017}. Even after the conclusion of the challenge, the dataset provided by the organizers continues to be widely used by research groups for their own developments and now serves as a standard reference in the CT community for reconstruction advances.

Our goal with the fastMRI challenge project was provide a similar stimulation to machine learning research in MR image reconstruction aimed at reducing MR examination times. In December of 2018, we released the first large-scale database of MRI scanner raw data from a clinical patient population~\cite{Zbontar2019,Knoll2019_fastMRI}. In the spirit of previous challenges organized by the ISMRM community~\cite{Grissom2017}, we then conducted a challenge to provide researchers in the field the opportunity to evaluate their methods in a large-scale, realistic setting with evaluation from clinical radiologists.  We also aimed to spark interest in radiology and biomedical imaging within the large machine learning and computer vision research community. In this article, we describe the design and the results of the challenge as well as the lessons we learned from its organization.

\section{Methods}
\subsection{Challenge design principles}
Our challenge was focused on accelerating MR image acquisitions. Two of the most influential developments in this arena during the last two decades have been parallel imaging ~\cite{Sodickson1997,Pruessmann1999,Griswold2005} and compressed sensing~\cite{Lustig2007}. Both of these approaches to rapid imaging are based on the principle of reducing the number of lines that are acquired in k-space, which reduces the scan time, and then exploiting redundancy in the measured data during the image reconstruction process. In parallel imaging, the redundancy arises from the simultaneous acquisition of MR signal with multiple receive coils; in compressed sensing, it derives from the observation that images are generally compressible. Machine learning approaches have generally adopted similar strategies for the acceleration of MRI, which set the main design criteria for our challenge.

We provided participants with sets of raw k-space data, and the goal of the challenge was to reconstruct images from these data. Since details about the dataset are reported in separate publications~\cite{Zbontar2019,Knoll2019_fastMRI}, in this article we restrict our description of the dataset only to information that is relevant to the design of the challenge. We provided data for a total of 1,594 consecutive clinical proton-density-weighted MRI acquisitions of the knee in the coronal plane, both with (COR PD FS) and without (COR PD) frequency-selective fat saturation. In addition to their different image contrast, these two types of acquisition also vary in signal to noise ratio (SNR) by approximately a factor of 4~\cite{Knoll2019}. Data were acquired on three clinical 3T systems (Siemens Magnetom Skyra, Prisma, and Biograph-mMR) and one clinical 1.5T system (Siemens Magnetom Aera) using clinical multi-channel receive coils. Curation of the dataset was part of a study approved by our local institutional review board (IRB). 

The selection of problems for the challenge was based on a three-way trade-off between a) providing a realistic scenario representative of actual clinical imaging exams, b) allowing fair and proper validation, and c) making the challenge practically and conceptually accessible for research groups outside the core field of MR image reconstruction. This led to the following design principles:

\begin{itemize}
    \item To make the image reconstruction problem realistic, we provided actual raw (complex valued) k-space data obtained directly from our MRI scanners.
    \item To reduce the complexity of the challenge, we restricted ourselves to standard Cartesian 2D Turbo Spin Echo sequences that are part of the routine clinical protocol at our institution.
    \item In order to provide clear ground truth against which to compare image reconstructions, we did not provide prospectively undersampled data. Fully-sampled k-space data were acquired for all exams in the data set, and undersampling was performed retrospectively, so that no differences in conditions (e.g., in motion state or scanner calibration) between fully-sampled and undersampled acquisitions would complicate image comparisons. 
    \item Since the goal of the challenge was to test reconstruction methods and not sampling trajectory design, we predefined the allowed undersampling patterns. We chose one-dimensional pseudo-random sampling in the phase encoding direction, with full sampling of a small central k-space region, as introduced in the context of compressed sensing~\cite{Lustig2007}.
    \item For multi-coil acquisitions, our ground truth reference was the root-sum-of-squares combination of the fully-sampled multi-channel data after inverse Fourier transform. While this is not the optimal coil combination method in terms of SNR~\cite{Roemer1990,Walsh2000}, it does not bias the ground truth towards any particular approach to the estimation of coil sensitivities. We also removed readout-direction oversampling by cropping the reconstructed images to the central $320 \times 320$ pixel region. For the single-coil case, which is uncommon in clinical practice but was included to provide a low barrier to entry for those not familiar with multi-coil data acquisitions, we simulated a physically feasible ground truth using a linear combination of individual coil signals as described in \cite{Tygert2018}.
    \item We used the structural similarity index (SSIM)~\cite{Wang2004} with respect to the fully sampled ground truth reference as an indicator of image quality. We calculated two other widely used quantitative metrics for our online leaderboard: pixelwise normalized root mean square error (NRMSE) and peak signal to noise ratio (PSNR). However, since all of these metrics provide limited insight into the diagnostic quality of medical images, we decided to use a ranking by an expert panel of musculoskeletal radiologists as the final metric to determine the winning entries in the challenge.
    \item We did not prohibit the use of additional non-fastMRI data in the development and training of the submissions. However, all participants who chose to use additional data were required to state this at the time of submission.
\end{itemize}

\subsection{Challenge tracks}
One of our design goals was to test the submissions in different operating modes defined by the level of acceleration. We also wanted to make the challenge interesting for research groups with a focus on MR image reconstruction as well as for groups based in machine learning, computer vision and image processing. We therefore decided to organize the challenge into multiple submission tracks.

Regarding the different levels of acceleration, the goal of the first scenario was to operate in a mode where we expected the reconstruction to be challenging, but where reconstructed images that might be acceptable for clinical diagnosis were likely to be feasible. Based on our previous experience with similar data~\cite{Hammernik2017,Knoll2019}, we chose an undersampling factor of R=4 for this scenario. The goal of the second scenario was to aim for a substantially higher acceleration than can be achieved with current reconstruction methods. We chose an undersampling factor of R=8 for this scenario. We did not expect to receive submissions with clinically acceptable image quality at this high level of acceleration. The goal for this scenario was to evaluate the performance of the submissions when they were pushed beyond reasonable limits, and to analyze failure modes.
   
In our experience, the steepest component of the learning curve for use of (Cartesian) MR data in image reconstruction relates to the proper handling of multi-channel raw k-space data of the sort required for parallel imaging. We therefore designed two additional tracks, which we termed the multi-coil and the single-coil track. For the multi-coil track, which was primarily aimed at research groups with a background in MR image reconstruction, we provided true multi-channel raw data from the MR scanners. Since most modern MR scans are performed using arrays of detector coils, this is a realistic scenario whose results are likely to be readily translatable to real-world imaging situations. For the single-coil track, we provided k-space data for which multi-channel information had been combined into a single channel that can be reconstructed with a simple inverse Fourier transform in the fully sampled case. However, instead of Fourier transforming previously-reconstructed images stored in DICOM format in an attempt to create k-space data (an approach sometimes observed in the literature, but not realistic or advisable for various reasons), we chose to retain the complex nature of the original data. A more detailed description of the channel combination we used is presented in~\cite{Tygert2018}. The single-coil track was primarily aimed at research groups from machine learning, computer vision and image processing, who might be interested in applying their expertise in medical imaging applications. In particular, after a simple inverse Fourier transform of the data, the single-coil track enabled easy use of methods that are entirely based on image postprocessing. While the removal of multi-channel information decreases the complexity of working with the data, it also increases the difficulty of the reconstruction problem at any given acceleration factor, since the resulting single-channel data has reduced redundancy and more limited information content than the original multi-channel data.  For the challenge phase, we therefore decided to limit ourselves to a single acceleration factor of R=4 in this track.

\subsection{Dataset split and leaderboard evaluation}
We partitioned our dataset into six subsets for the individual tracks and the different phases of the challenge: training, validation, multi-coil test, single-coil test, multi-coil challenge, or single-coil challenge. Data from individual patient cases were randomly assigned to the individual subsets. The number of cases and the total number of slices are shown in Table~\ref{tab:dataset_stats}. The dataset was made publicly available at https://fastmri.med.nyu.edu/.

\begin{table*}[ht!]
\caption{Overview of the dataset that was provided for the fastMRI challenge.}
\label{tab:dataset_stats}
\centering
\begin{tabular}{ccccc}
\hline
 & \multicolumn{2}{c}{Cases} & \multicolumn{2}{c}{Slices} \\ 
& Multi-coil & Single-coil & Multi-coil & Single-coil \\
\hline
training & 973 & 973 & 34,742 & 34,742 \\
validation & 199 & 199 & 7,135 & 7,135 \\
test & 118 & 108 & 4,092 & 3,903 \\
challenge & 104 & 92 & 3,810 & 3,305 \\
\hline
\end{tabular} 
\end{table*}

We provided fully sampled k-space data and corresponding ground truth image reconstructions for the training and validation subsets, which could be used by the participants to develop and train their machine learning models and to determine any hyperparameters. In order to make most efficient use of our available data, we used the same training and validation cases for the multi-coil and single-coil tracks.

\begin{figure*}[ht!]
    \centering
    \begin{subfigure}[b]{\textwidth}
        \centering
	    \includegraphics[width=0.5\textwidth]{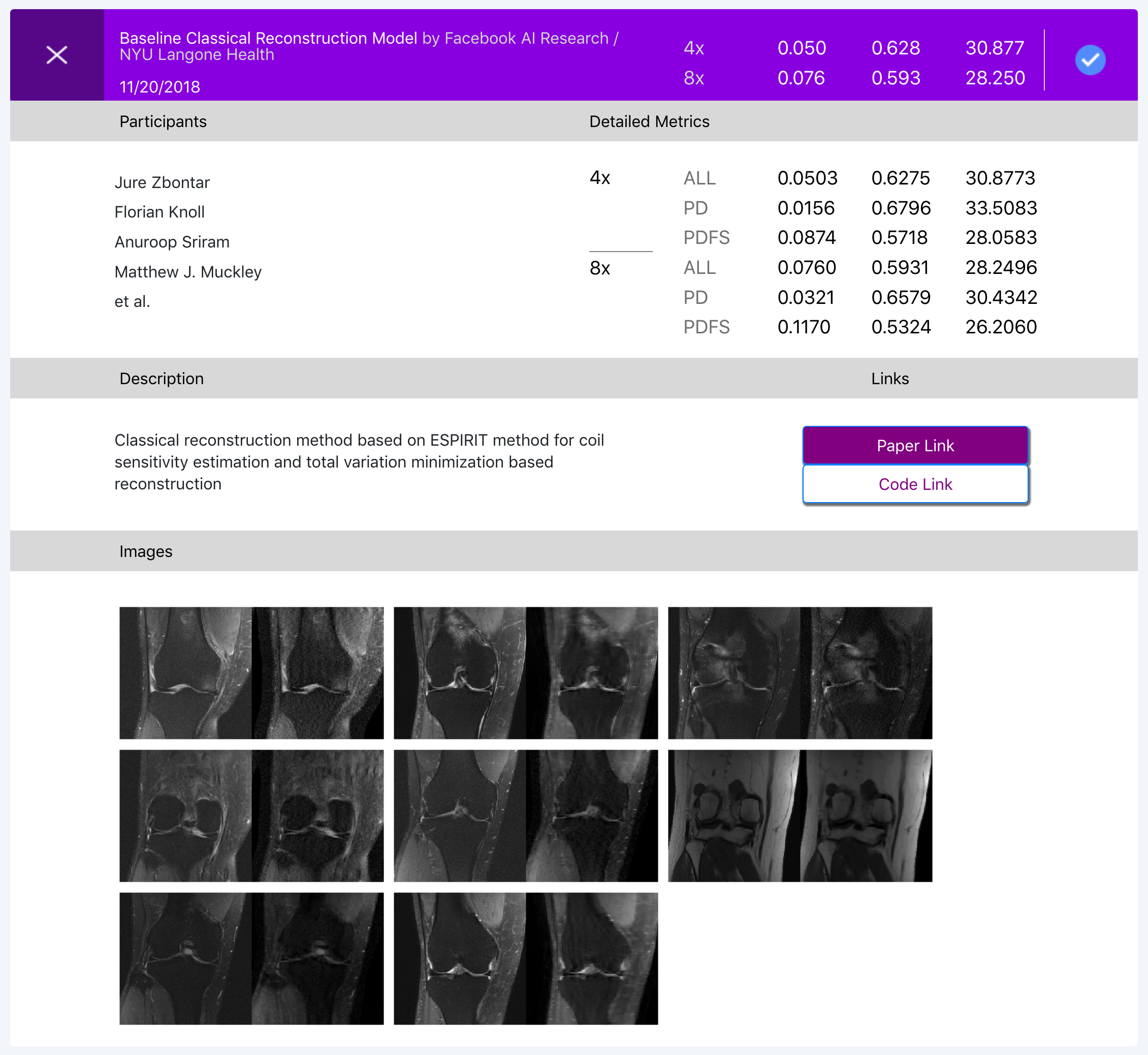}
	    \caption{Overview of Facebook AI Research and NYU baseline entries for the multi-coil test-set leaderboard. (Baseline entries with modest performance were provided for reference.)  Three quantitative metrics are provided for R=4 and R=8 for both image contrasts, along with a selection of reconstructed images. A short description of the reconstruction approach used, together with  links to a corresponding paper or code repository, are also shown if the submitting groups provide this information.}
	     \label{fig:leaderboard_test}
  	\end{subfigure}
  	\begin{subfigure}[b]{\textwidth}
  	    \centering
	    \includegraphics[width=0.5\textwidth]{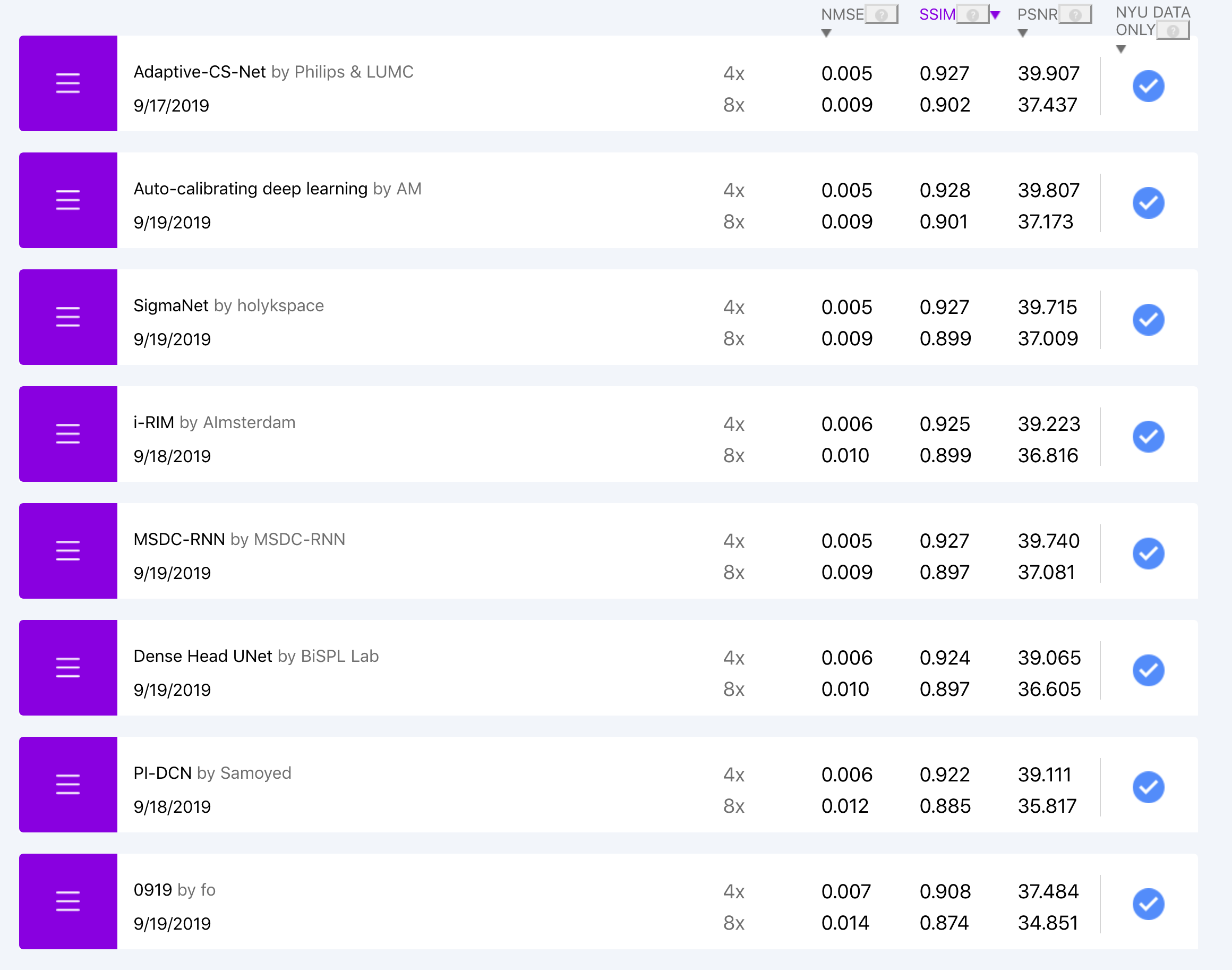}
	     \caption{Challenge leaderboard showing the 8 submissions for the multi-coil track.}
	     \label{fig:leaderboard_challenge}
 	\end{subfigure}
   \caption{Online leaderboard at the completion of the challenge (December 2019).}
\end{figure*}

For the test set, we provided different subsets of undersampled k-space data for the single and the multi-channel tracks. Participants could upload their reconstruction results to our public leaderboard at http://fastmri.org/. We then calculated NRMSE, PSNR and SSIM and evaluated performance on the complete test dataset as well as individual errors for the two image contrasts (with and without fat suppression). Participants could submit to each track leaderboard once a day before the submission deadline. Submissions were ranked by SSIM of R=8 undersampling. A screenshot of baseline entries for the multi-coil track, provided by Facebook AI Research and NYU for reference, is shown in Figure~\ref{fig:leaderboard_test}. In addition to the quantitative scores, we also show a selection of reconstructed images on the leaderboard. 

The challenge dataset was released on September 5th, 2019 (see below for a description of the timeline), and the submission window was then open for 14 days. The evaluation and the structure of the leaderboard for the challenge phase were identical to those for the test phase, but each team could only make one challenge submission, and challenge results were made available only after the submission window was closed. A screenshot of the leaderboard for the multi-channel track of the completed challenge is shown in Figure~\ref{fig:leaderboard_challenge}.

\subsection{Challenge timeline and design of the evaluation}
The challenge consisted of multiple phases according the following timeline:

\begin{itemize}
\item November 26, 2018: Release of the training and validation sections of the fastMRI dataset~\cite{Zbontar2019,Knoll2019_fastMRI}.

\item June 5, 2019: Official announcement of the challenge and release of the test set. The test set leaderboard was open for submission at this stage.

\item September 5 to 19, 2019: Release of the challenge dataset and challenge submission window.

\item September 19, 2019: Quantitative evaluation of the challenge submissions. We selected the top 4 submissions with highest SSIM on the challenge dataset from each track for the second phase of evaluation by a panel of radiologists. At this stage, we also asked all participants to provide an abstract for the 2019 Medical Imaging Meets NeurIPS workshop\footnote{https://sites.google.com/view/med-neurips-2019}.

\item September 20 to October 10, 2019: Radiologist evaluation phase. We sent 5 randomly selected cases (with and without fat suppression) from the top 4 submissions in each track plus the corresponding ground truth reconstructions to our panel of seven radiologists from multiple institutions, including NYU Langone Health, Cleveland Clinic, University of California San Diego, University of Wisconsin and Stanford University. Each radiologist looked at a total of 1840 images. We asked the panel to rank the submissions in terms of the overall image quality to select one winner in each track. We then averaged the rankings of the radiologists to determine the winners. In addition, we asked the radiologists to score each submission on a 4-point scale (1 is best and 4 is worst) for the following criteria: Presence of artifacts, image sharpness, perceived contrast-to-noise ratio and diagnostic confidence. This rating was performed to obtain additional meta-information about the readers' preferences, and to give the radiologists some suggestions on which to base their ranking. Subcriterion rankings were not directly used to determine the winners of the challenge. At the end of this stage, we notified the winners of the three tracks and shared their abstracts and identity with the organizers of the 2019 Medical Imaging Meets NeurIPS workshop.

\item December 1, 2019: Publication of the challenge leaderboard with the results of the quantitative evaluation.

\item December 14, 2019: Official announcement of the winners of the three tracks at the 2019 Medical Imaging Meets NeurIPS workshop, with oral presentation by the three winning teams.
\end{itemize}

\section{Results}

\subsection{Overview of submissions}
We received a total of 33 challenge submissions. 8 groups submitted to the multi-coil track, each with submissions for both the R=4 and the R=8 tracks. At the time of this writing, two of the submitting groups have published manuscripts on their approach, in addition to their NeurIPS abstracts: $Sigma-Net$ from team holykspace~\cite{schlemper2019net} and iRim from team AImsterdam~\cite{putzky2019irim}. 17 groups submitted to the single coil R=4 track. 6 out of the 8 groups who submitted to the multi-coil track also submitted to the single-coil track. We did not require that the groups publicly disclose their names or affiliations at the submission stage. This was only required for the winners of each track. For the test-set leaderboard during the full duration of the challenge, we received more than 25 submissions for the multi-coil track and more than 70 submissions for the single coil track. All submissions used exclusively fastMRI data in the training of their approach for both challenge and public test submissions. We encouraged all participants to provide open source code together with their submissions, and 3 groups provided links to their open source code repositories.

\subsection{Analysis of results}

\begin{figure*}
    \centering
    \begin{subfigure}[b]{\textwidth}
        \centering
        \includegraphics[width=1\textwidth]{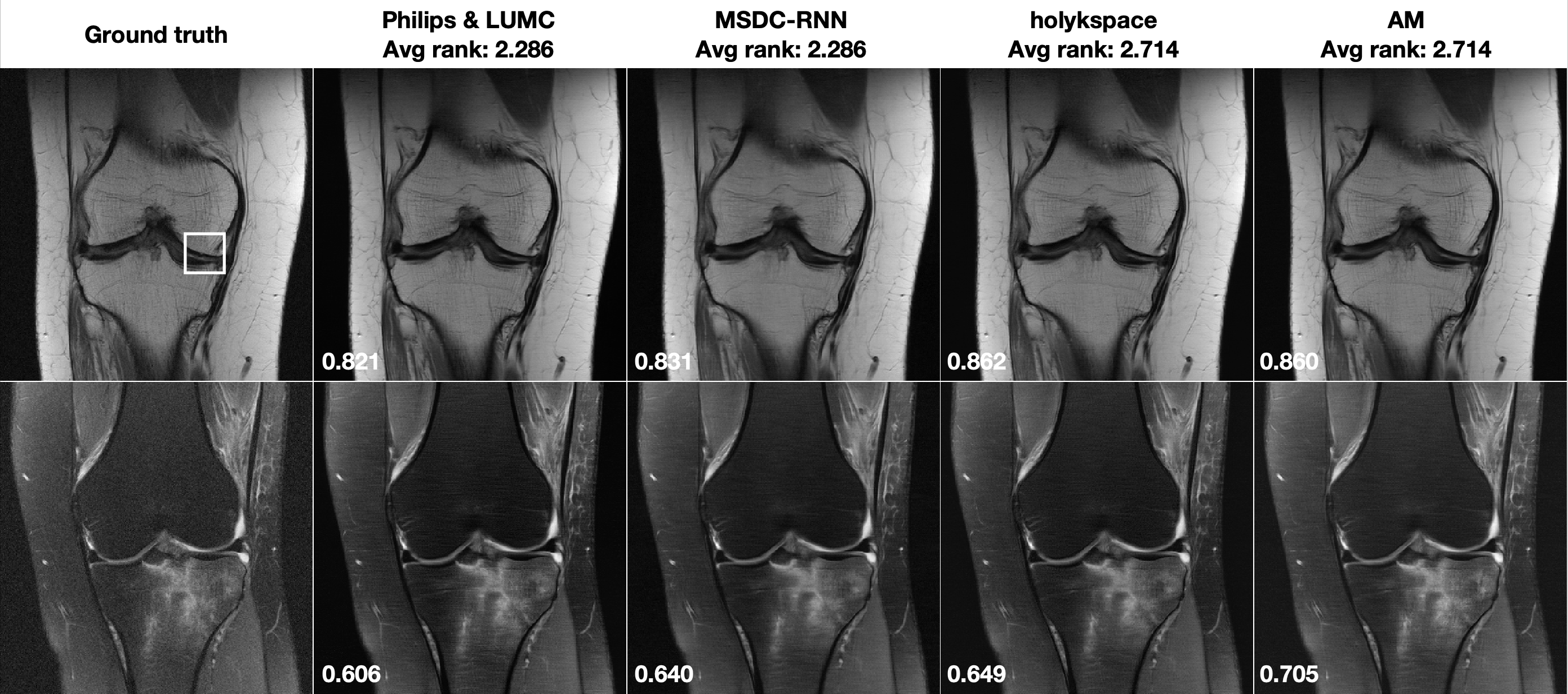}
        \caption{Top row: Results for one slice from an acquisition without fat suppression. This case shows subtle pathology in the ROI indicated by a white rectangle in the ground truth image. Bottom row: One slice from an acquisition with fat suppression.}
    \end{subfigure}
    \begin{subfigure}[b]{\textwidth}
        \centering
        \includegraphics[width=1\textwidth]{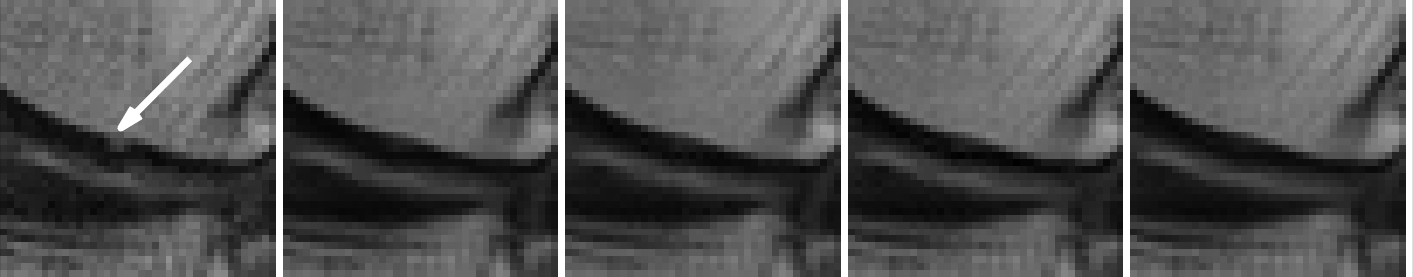}
        \caption{Zoomed view of the ROI that shows a subchondral osteophyte (highlighted by a white arrow in the ground truth reconstruction). This pathology is not visible in any of the accelerated reconstructions.}
    \end{subfigure}
    \caption{Multi-Coil R=4 track results: Selected results from the top 4 submissions in each track, for both image contrasts. The submissions are ordered from left to right based on the average of radiologists' rankings. SSIM to the ground truth for this particular slice is displayed in the bottom-left corner of each image.}
    \label{fig:results_mc4x}
\end{figure*}

Figure~\ref{fig:results_mc4x} shows selected results from the Multi-Coil R=4 track, for one particular slice with and one slice without fat suppression. Both of these cases were obtained from 1.5T systems (Siemens Magnetom Aera). The top 4 submissions that were evaluated by the radiologists are ordered from left to right based on the radiologists rankings, next to the ground truth on the far left. The average rank of the 7 radiologists is displayed on top of each submission. The SSIM to the ground truth for each particular slice is shown in the bottom left of the plots. The case in the top row shows a subtle subchondral osteophyte, which was not visible in the accelerated reconstructions.

\begin{figure*}
    \centering
    \begin{subfigure}[b]{\textwidth}
        \centering
        \includegraphics[width=1\textwidth]{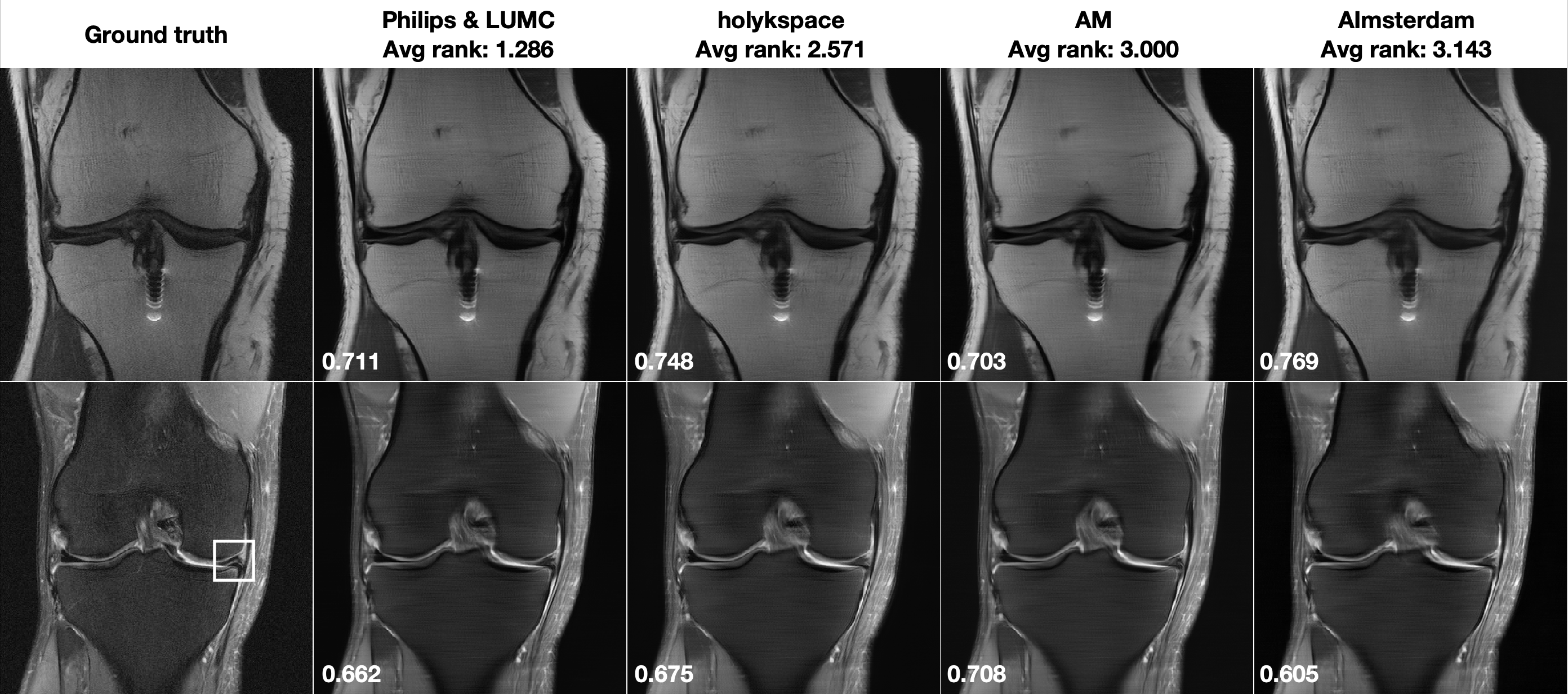}
        \caption{Top row: Results for one slice from an acquisition without fat suppression. This case shows shows moderate artifact from a metal implant. Bottom row: One slice from an acquisition with fat suppression. This case shows shows a meniscal tear in the ROI indicated by a white rectangle in the ground truth image.}
	\end{subfigure}
	\begin{subfigure}[b]{\textwidth}
        \centering
        \includegraphics[width=1\textwidth]{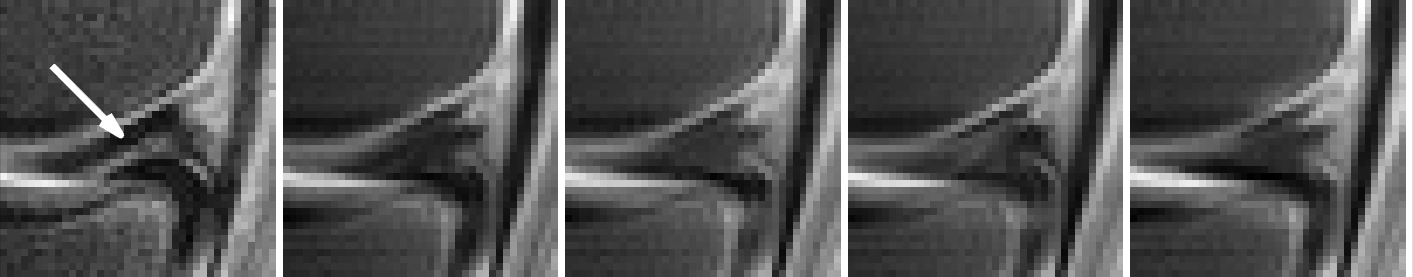}
        \caption{Zoomed view of the ROI that shows a meniscal tear (highlighted by a white arrow in the ground truth reconstruction). This pathology is not well seen in any of the accelerated reconstructions.}
	\end{subfigure}
	\caption{Multi-Coil R=8 track results: Selected results from the top 4 submissions in each track. The submissions are ordered from left to right based on the average of radiologists' rankings. SSIM to the ground truth for this particular slice is displayed in the bottom-left corner of each image.}
    \label{fig:results_mc8x}
\end{figure*}

Figure~\ref{fig:results_mc8x} shows results for the Multi-Coil R=8 track. The case in the top row shows moderate artifact from a metal implant and was obtained on a 1.5 system (Siemens Magnetom Aera). None of the submissions was negatively affected by this irregularity. The case in the bottom row shows a meniscal tear. It was acquired on a 3T system (Siemens Magnetom Prisma). This pathology was not visible in the accelerated reconstructions.

\begin{figure*}
    \centering
	\includegraphics[width=1\textwidth]{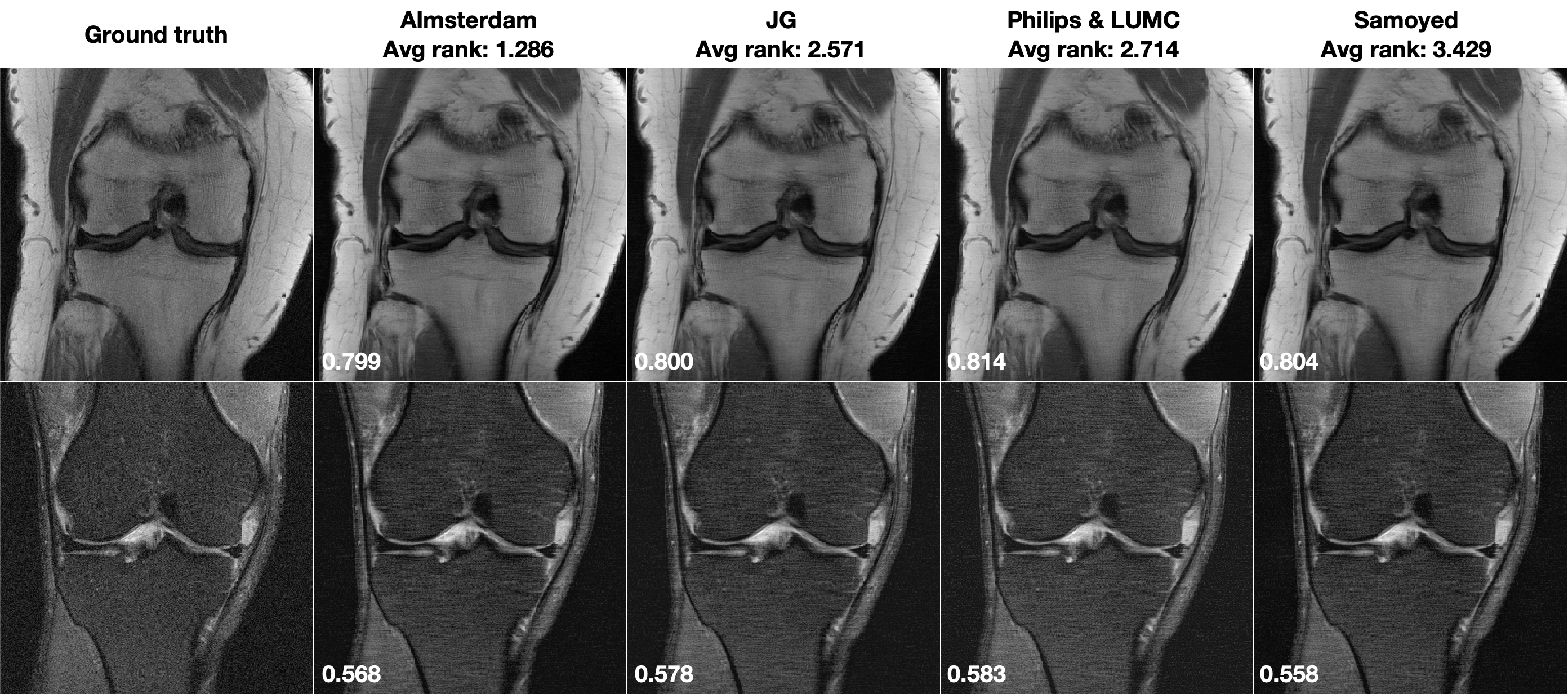}
	\caption{Single-Coil R=4 track results: Selected results from the top 4 submissions for each track.}
	\label{fig:results_sc4x}
\end{figure*}

Figure~\ref{fig:results_sc4x} shows results for the Single-Coil R=4 track. Both of these cases were obtained from 3T systems (Siemens Magnetom Skyra).

\begin{figure}
    \centering
    \includegraphics[width=\columnwidth]{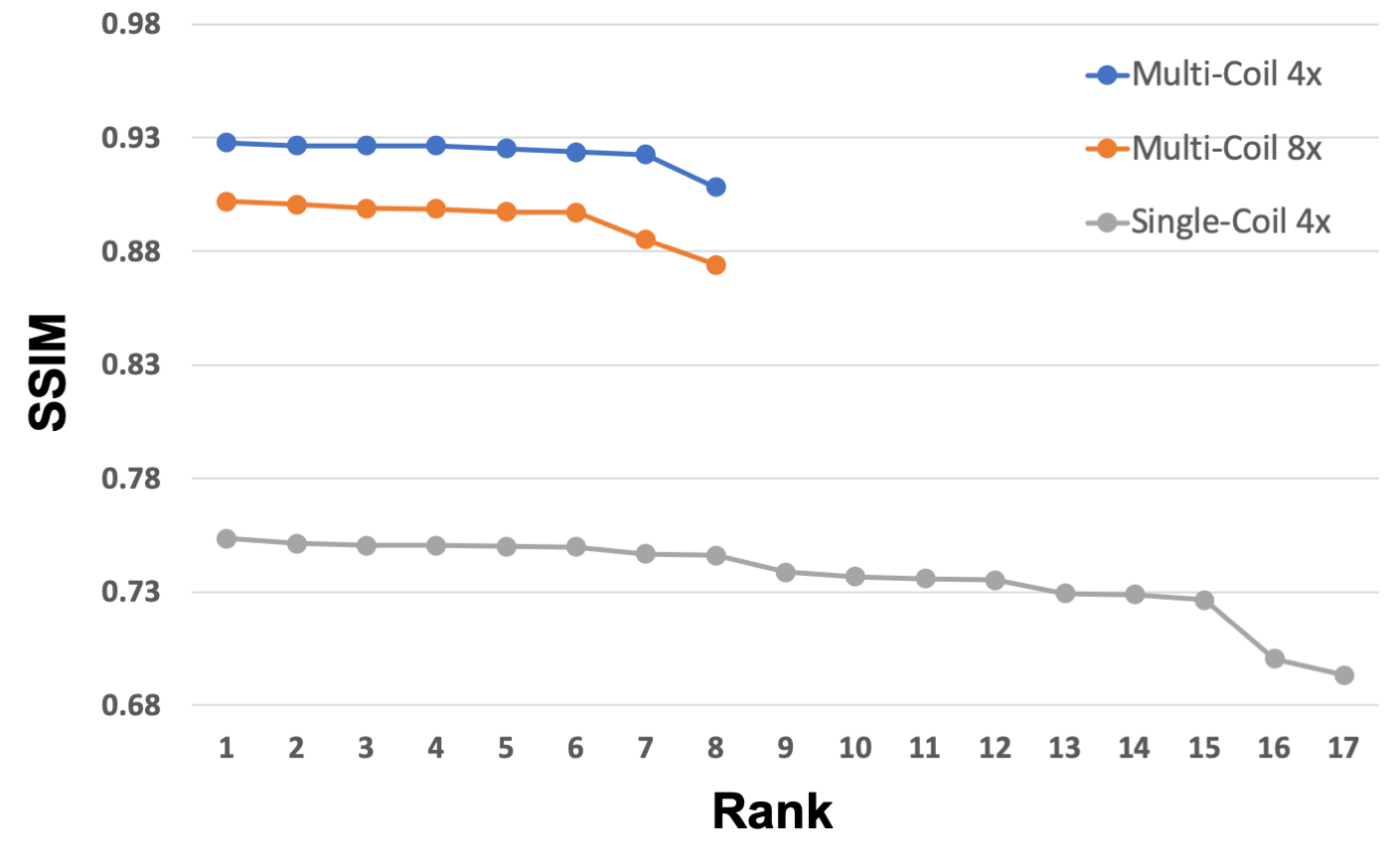}
	\caption{SSIM scores of the challenge submissions for each track. As expected, there is a substantial difference in overall SSIM values between the multi-coil and the single-coil tracks.}
    \label{fig:ssim_challenge}
\end{figure}

The SSIM scores of the challenge submissions are shown in Figure~\ref{fig:ssim_challenge}. As expected, there is a substantial difference in overall SSIM values between the multi-coil and the single-coil tracks. The average SSIM of all submissions was 0.924, 0.895 and 0.707 for the multi-coil R=4, multi-coil R=8 and single-coil R=4 tracks, respectively. Even the lowest-ranking multi-coil R=8 submission (SSIM=0.874) significantly outperformed the highest-ranking single-coil R=4 submission (SSIM=0.754).

\begin{figure}
    \centering
    \includegraphics[width=\columnwidth]{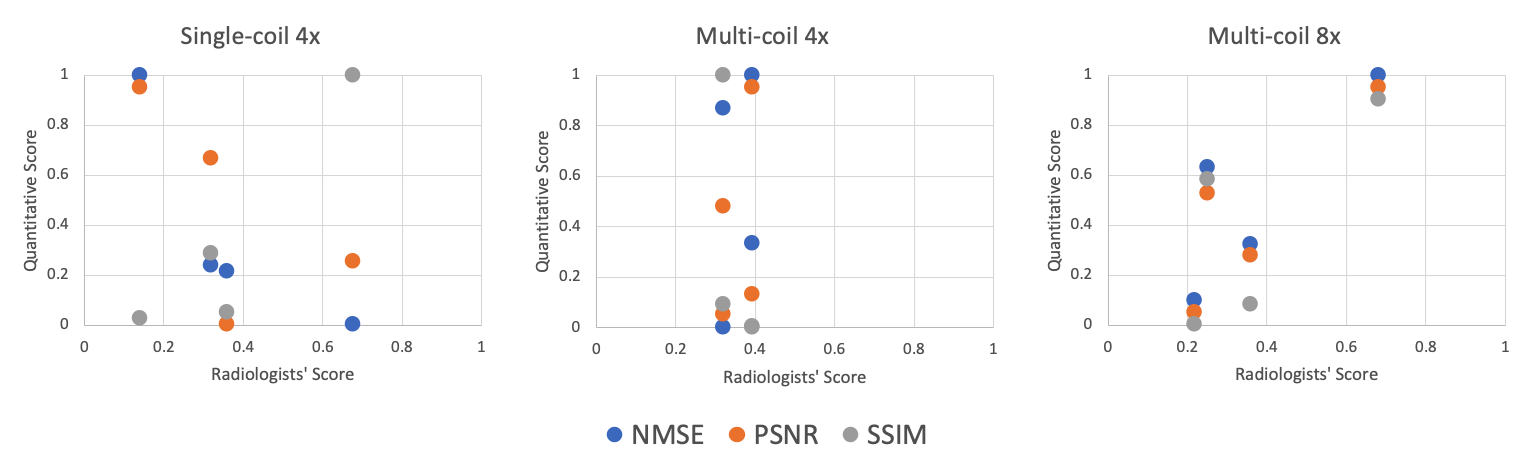}
	\caption{Scatterplots of the NMSE, PSNR, and SSIM scores (normalized so that the best score corresponds to a value of 1, for convenient visualization) versus the average radiologists' score based on ranking (1 is best) for the top 4 submissions in all three submission tracks.}
    \label{fig:ssim_ranks_correlation}
\end{figure}

Table~\ref{tab:radiologist_ranking} shows the average radiologist rankings as well as the overall SSIM, RMSE and PSNR values for the full challenge dataset for the top 4 submissions of each track. Figure~\ref{fig:ssim_ranks_correlation} shows corresponding scatterplots after normalization of the scores (1 is best). In the case of multi-coil R=8, the highest ranked submission was also the one that had the highest SSIM, RMSE and PSNR values. For and single-coil R=4, only SSIM showed a similar trend as the radiologists scores, while the other two metrics showed almost opposite trends. For the multi-coil R=4 track, the top 4 submissions were very close together with all metrics. The differences in SSIM between the submissions were less than 1\%.

 \begin{table*}
\caption{Average radiologist rankings and corresponding SSIM, RMSE and PSNR scores for the full challenge dataset for the top 4 submissions of each track.}
\label{tab:radiologist_ranking}
\begin{subtable}{1\textwidth}
    \centering
    \caption{Multi-Coil R=4.}
    \begin{tabular}{cccccc}
    Team name            & Rank     & Avg radiologist rank  & SSIM      & RMSE      &    PSNR\\ 
    \hline
    Philips \& LUMC      & 1(tie)   & 2.285                 & 0.927     & 0.005     & 39.907 \\ 
    MSDC-RNN             & 1(tie)   & 2.285                 & 0.927     & 0.005     & 39.740 \\
    holykspace           & 3(tie)   & 2.714                 & 0.927     & 0.005     & 39.715 \\
    AM                   & 3(tie)   & 2.714                 & 0.928     & 0.005     & 39.807\\
    \hspace{1pt}
    \end{tabular} 
\end{subtable}
\begin{subtable}{1\textwidth}
    \centering
    \caption{Multi-Coil R=8.}
    \begin{tabular}{cccccc}
    Team name              & Rank     & Avg radiologist rank  & SSIM    & RMSE      &    PSNR\\ \\ 
    \hline
    Philips \& LUMC        & 1        & 1.286                 & 0.901   & 0.0086    & 37.437\\ 
    holykspace             & 2        & 2.571                 & 0.899   & 0.0092    & 37.009\\ 
    AM                     & 3        & 3.000                 & 0.901   & 0.0089    & 37.173\\ 
    AImsterdam             & 4        & 3.143                 & 0.898   & 0.0096    & 36.816\\ 
    \hspace{1pt}
    \end{tabular} 
\end{subtable}
\begin{subtable}{1\textwidth}
    \centering
    \caption{Single-Coil R=4.}
    \begin{tabular}{cccccc}
    Team name              & Rank     & Avg radiologist rank  & SSIM    & RMSE      &    PSNR\\ \\ 
    \hline
    AImsterdam            & 1        & 1.286                 & 0.754 & 0.031    & 32.549\\ 
    JG                    & 2        & 2.571                 & 0.750 & 0.031    & 32.476\\ 
    Philips \& LUMC       & 3        & 2.714                 & 0.751 & 0.030    & 32.666\\ 
    Samoyed               & 4        & 3.428                 & 0.751 & 0.029    & 32.761\\ 
    \hspace{1pt}
    \end{tabular} 
\end{subtable}
\end{table*}


\begin{figure}
    \centering
    \includegraphics[width=\columnwidth]{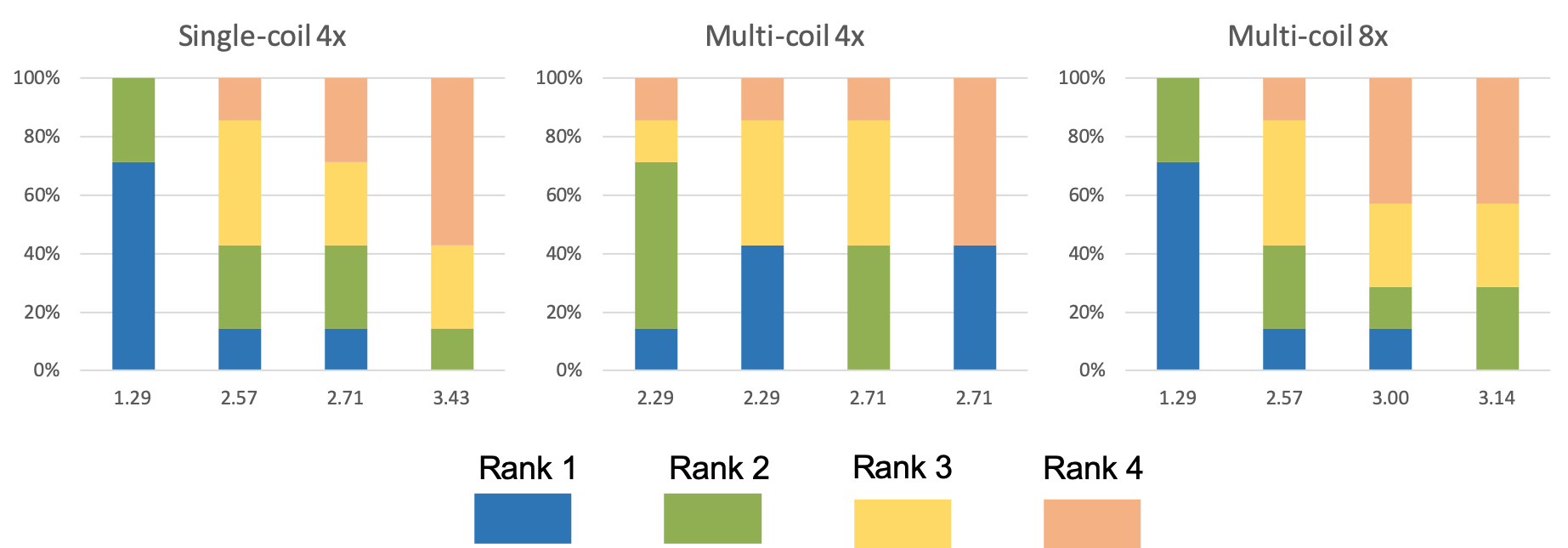}
	\caption{Individual rankings by the 7 radiologists for the top 4 submissions in all three submission tracks.}
    \label{fig:radiologists_rank_correlations}
\end{figure}

Additional insight into the radiologists' ratings is provided by Figure~\ref{fig:radiologists_rank_correlations}, which shows the individual rankings by the 7 radiologists for the top 4 submissions in all three submission tracks. For multi-coil R=8 and single-coil R=4 tracks, the radiologists had a strong preference for a single submission. The highest-rated submission in each of these tracks was ranked first by 5 radiologists, and ranked second by the remaining 2 radiologists. The results are substantially less consistent for the multi-coil R=4 track. The two highest-rated submissions each were also ranked worst by one reader. The lowest-rated submission was actually ranked best by 3 out of 7 radiologists, and ranked worst by the remaining 4.

Table~\ref{tab:radiologist_ranking_categories} shows the average scores for the individual categories that the radiologists were asked to rate for the top 4 submissions in each track: Artifacts, sharpness, perceived contrast-to-noise ratio and and diagnostic confidence, using a 4 point scale, where 1 is best and 4 is worst. These categories were intended as guidelines for radiologist ranking, not as strict criteria. However, by and large the radiologists chose to rank the submissions based on the sum of their scores in the different categories. For the multi-coil R=8 track, all radiologists ranked the submissions strictly based on their scores. 5 out of 7 radiologists for the multi-coil R=4 track and 6 out of 7 radiologists for the single-coil R=4 track ranked submissions strictly based on their scores. Even for the cases where the ranking deviated from the scores, the top-ranked submission always had the best scores as well.

\begin{table*}
\caption{Average ratings for the individual categories that the 7 radiologists were asked to rate, for the top 4 submissions from each track.  Ratings followed a 4 point scale, where 1 is best and 4 is worst.}
\label{tab:radiologist_ranking_categories}
\begin{subtable}{1\textwidth}
    \centering
    \caption{Multi-Coil R=4 track. 5 out of 7 radiologists based their ratings strictly on their scores for this track. For the remaining 2 radiologists, the top-ranked submission also had the best overall score.}
    \begin{tabular}{ccccc}
    Team name        & Artifacts    & Sharpness    & Contrast to noise     & Diagnostic Confidence \\ 
    \hline
    Philips \& LUMC	& 2.714	& 2.286	& 2.286	& 2.000	\\
    MSDC-RNN	    & 2.571	& 2.286	& 2.429	& 1.857	\\
    holykspace	    & 2.000	& 3.000	& 2.714	& 1.857 \\
    AM      	    & 2.000	& 3.000	& 2.000	& 2.000	\\
    \hspace{1pt}
    \end{tabular} 
\end{subtable}
\begin{subtable}{1\textwidth}
    \centering
    \caption{Multi-Coil R=8 track. All radiologists based their ratings strictly on their scores for this track.}
    \begin{tabular}{ccccc}
    Team name       & Artifacts       & Sharpness    & Contrast to noise     & Diagnostic Confidence  \\ 
    \hline
    Philips \& LUMC	& 1.714	& 2.286 & 2.286 & 2.286 \\
    holykspace	    & 2.143 & 3.143 & 2.286 & 2.857 \\
    AM 	            & 1.857 & 3.286 & 2.429 & 3.143  \\
    AImsterdam      & 2.714	& 2.857	& 3.000	& 3.143	\\
    \hspace{1pt}
    \end{tabular} 
\end{subtable}
\begin{subtable}{1\textwidth}
    \centering
    \caption{Single-Coil R=4 track. 6 out of 7 radiologists based their ratings strictly on their scores for this track. For the remaining radiologist, the top-ranked submission also had the best overall score.}
    \begin{tabular}{ccccc}
    Team name               & Artifacts      & Sharpness       & Contrast to noise  & Diagnostic Confidence \\ 
    \hline
    AImsterdam	            & 2.429	         & 2.286	       & 2.143	            & 2.286	                   \\
    JG	                    & 3.000	         & 2.714	       & 2.429	            & 2.571	                    \\
    Philips \& LUMC	        & 2.714	         & 3.000	       & 2.714	            & 2.857	                    \\
    Samoyed	                & 3.143	         & 3.286	       & 3.000	            & 3.286	                    \\
    \hspace{1pt}
    \end{tabular} 
\end{subtable}
\end{table*}

\section{Discussion}

\subsection{Limits of the challenge design}
One of our most consequential decisions in terms of challenge design was to not generate any systematic differences between training, validation, test and challenge sets. All of these datasets were randomly selected from the same superset of data, and all consisted of coronal knee data from a limited set of MR scanners from a single vendor. This design substantially limits insight into robustness and generalization. It is possible to subsequently perform a more targeted analysis by, for example, only using a subset of the training data from one of the two contrasts or one field strength (1.5T or 3T) for training and validation, and the other set of data for testing. Given the importance of multi-coil data in the overall performance of the submissions, the challenge also didn't include substantial variations of receive coil geometries. All coil arrays were standard knee coil configurations from a single vendor, with the same number of receive channels (15).

In addition to maintaining homogeneity of the data on a technical level, we also decided not to perform curation of the data in terms of anatomical or pathological variations. An interesting follow-up challenge would involve separating pathological from non-pathological cases and using only one of these individual subsets for training and validation, and the second subset for testing. Aside from pathology, similar experiments could be performed by grouping subsets of data based on age, height, weight, body mass index or gender.

In terms of the evaluation, a substantial limitation was that we did not evaluate diagnostic interchangeability of accelerated reconstructions with the fully sampled ground truth reconstruction, but only asked radiologists to rate image submissions by image quality on a subjective level. A more detailed discussion of this limitation is provided in the next section.

\subsection{Analysis of the submissions and results}
The quantitative SSIM values (Figure~\ref{fig:ssim_challenge}) provide several interesting insights. First, the differences in SSIM values between submissions from the top teams are almost negligible. In each of the three tracks, the difference between the first- and the fourth-ranked entry was less than 1\%. For the multi-coil R=4 track, the radiologists' scoring showed a similar trend. Both the top two and the bottom two submissions were tied in the ranking. However, since all participants decided to use only NYU-provided training data, no conclusions can be drawn about potential improvements by using additional training data, either by expanding the dataset with additional knee data, or by using synthetic data and transfer learning.

It is often pointed out in the medical imaging community that quantitative metrics like RMSE, PSNR and SSIM are poor metrics to evaluate the quality of medical images. In our challenge, juxtaposition of the radiologists' scores with SSIM values ~(Figure~\ref{fig:ssim_ranks_correlation}) shows that for the two tracks where the radiologists picked a clear winner (multi-coil R=8 and single-coil R=4), the winner was also the submission with the highest SSIM value. RMSE and PSNR were aligned with this trend for multi-coil R=8, but for single-coil R=4, RMSE actually resulted in the opposite ranking order from that selected by the radiologists. While none of the metrics in any track resulted in the same rank order as the radiologists, it is important to remember that the quantitative values were very closely spaced. It is interesting that for the track where the SSIM values of the top 4 submissions were essentially identical (multi-coil R=4), the individual radiologists also had substantial disagreement in their preference (Figure~\ref{fig:radiologists_rank_correlations}). The lowest-ranked submission was actually ranked best by 3 out of 7 radiologists, and ranked worst by the remaining 4. This indicates that for the multi-coil R=4 track, the submissions were most likely identical in terms of image quality, as correctly predicted by SSIM, and the ranking was determined by individual preferences for image quality by the radiologists. This means that in our challenge, SSIM actually did provide estimates of image quality that were consistent with the preferences of radiologists. Our results also suggest that radiologists' evaluations must be carried out at the level of diagnostic interpretation to allow their domain knowledge to provide substantial additional information.

While we knew that there would be a difference in performance between the single-coil and the multi-coil tracks, we were surprised by the degree of difference actually observed. From a linear algebraic point of view, the underlying problems in the different tracks are substantially different. The undersampled single-coil reconstruction problem is an undetermined system in which data acquisition violates the Shannon/Nyquist sampling theorem, and a solution can only be obtained by introducing prior knowledge and performing incoherent sampling, on the model of compressed sensing~\cite{Lustig2007}. By contrast, for the multi-coil problem, even at R=8 acceleration, the number of receive channels (15) is still higher than the undersampling factor. The underlying problem involves an overdetermined system. However, the problem is ill-posed because the individual coil elements do not provide independent information, and prior knowledge is also needed to constrain the solution. While this may raise the question of whether fundamentally different approaches should be developed for the two scenarios, the results of the challenge indicate otherwise. Six out of eight participants in the multi-coil track also submitted to the single-coil track. Team holykspace~\cite{schlemper2019net} and team AM used a dedicated approach for multi-coil data. Team AM explicitly estimated coil sensitivity maps using Espirit~\cite{Uecker2014a} and used a nullspace constraint on the fully-sampled center of k-space that is used to estimate the coil sensitivities. Team holykspace learned the implicit weighting of the individual coils. In contrast, the remaining groups used essentially the same core method for both tracks, and only fine-tuned and re-trained for the different tracks. Also, the top three submissions in the single-coil track were from the same groups that submitted to the multi-coil track. As expected, the number of submissions for the single-coil track was substantially higher than for the multi-coil track, most likely due to the shallower learning curve and greater ease of use of the single-coil data. However, the results from the challenge show that in order to achieve the best possible image quality for accelerated MR scans, it is essential to take the multi-channel nature of MR acquisitions into account. Therefore, we plan to limit ourselves to multi-coil tracks for future iterations of our challenge.

The inability to correctly identify subtle pathology, even in the multi-coil R=4 results in Figure~\ref{fig:results_mc4x}, must be considered in the light of clinical adoption. However, loss of low-contrast fine details is not necessarily a particular culprit of machine-learning-based reconstruction methods. It is entirely possible that this pathology would have been lost with any reconstruction approach at this level of acceleration. On the other hand, a common fear about machine learning reconstruction methods is that they react very unpredictably and unstably for cases that show severe abnormalities or deviations from normal anatomy. In our challenge, none of submissions showed any kind of deterioration for the case with the severe image artifact due to the metallic implant in the Multi-Coil R=8 track (Figure~\ref{fig:results_mc8x}). Separate dedicated studies will be required to investigate this effect, but this result is still encouraging from the point of view of robustness for clinical translation.

All submissions used a supervised learning approach with deep Neural Networks. While it is tempting to conclude that a similar paradigm shift towards deep learning has occurred for MR image reconstruction as in the ImageNet challenge~\cite{Russakovsky2015} for computer vision, in our opinion the results of our challenge do not allow us to draw that conclusion. First, the total number of submissions for our challenge was substantially smaller than for ILSVRC, and it was only the first time the challenge was held. Second, as described above, the design of the challenge essentially guaranteed that (supervised) machine learning methods would have strong performance on the challenge dataset. Third, in contrast to purely data-driven end-to-end learning in the true spirit of deep learning~\cite{LeCun2015}, the winners of all three tracks chose approaches that used a combination of a learned prior and a data-fidelity term that encodes information about the MR physics of the acquisition, in line with approaches that can be seen as neural network extensions of classic iterative image reconstruction methods~\cite{Hammernik2017,Schlemper2018,Aggarwal2018a}. Finally, even though radiologist ratings were ultimately the deciding factor that determined the winners, and while they did have the fully sampled ground truth available as a reference, their ratings were essentially based on subjective impression of image quality and not on diagnostic equivalency. The translation of machine learning for reconstruction of accelerated MRI scans in routine clinical practice remains an open question for future research and development.

\section*{Acknowledgements}
We first would like to thank all participants of the challenge. We thank the radiologists who provided the scoring for the second evaluation phase: Drs. Christine Chung and Mini Pathria of UCSD, Dr. Michael Tuite of University of Wisconsin, Dr. Christopher Beaulieu of Stanford, Drs. Naveen Subhas and Hakan Ilaslan of the Cleveland Clinic, and Dr. David Rubin of NYU Langone Health. We thank our external advisors for the organization of the challenge: Dr. Daniel Rueckert of Imperial College London, Dr. Jonathan Tamir of University of Texas at Austin, Dr. Joseph Cheng of Apple AI research and Dr. Frank Ong of Stanford. We also thank our colleagues Mark Tygert, Michal Drozdzal, Adriana Romero, Pascal Vincent, Erich Owens, Krzysztof Geras, Patricia Johnson, Mary Bruno, Jakob Asslaender, Yvonne Lui, Zhengnan Huang and Ruben Stern for their insights and feedback. We acknowledge grant support from the National Institutes of Health under grants NIH R01EB024532 and NIH P41EB017183.

\bibliographystyle{IEEEbib}
\bibliography{references}

\end{document}